\documentclass[conference]{IEEEtran}
\IEEEoverridecommandlockouts

\ifCLASSINFOpdf

\else

\fi

\usepackage[cmex10]{amsmath}

	\usepackage{mathtools}							
	\usepackage{amssymb,mathrsfs}				
	\usepackage{algorithmic}						
	\usepackage{algorithm}							
	\usepackage{calc}										
	\usepackage{relsize}								
	\usepackage{framed} 
	\usepackage{nomencl}
	\usepackage{subcaption}
	\usepackage[abbreviations = true]{siunitx}
	\DeclareSIUnit\year{yr}
	\DeclareSIUnit\dollar{\$}
	\DeclareSIUnit \VA { VA }
	\DeclareSIUnit \pu { p.u. }
	\DeclareSIUnit \MWh { MWh }
	\usepackage{longtable}							
	\usepackage{multirow}								
	\usepackage{array}									
	\usepackage{booktabs}
	\usepackage{tikz}
	\usetikzlibrary{calc}
	
	\usepackage{pgfplots}
	\usepackage{pstricks}
	\usepackage{fancyhdr}

\begin{document}

\title{Sample-Derived Disjunctive Rules for Secure Power System Operation \vspace{-0.5em}}


\author{
\IEEEauthorblockN{Jochen L. Cremer, Ioannis Konstantelos, Goran Strbac}
\IEEEauthorblockA{Department of Electrical and Electronic Engineering \\ Imperial College London  \\ London, United Kingdom\\
\{j.cremer16, i.konstantelos, g.strbac\}@imperial.ac.uk}\vspace{-3em}
\and
\IEEEauthorblockN{Simon H. Tindemans}
\IEEEauthorblockA{Department of Electrical Sustainable Energy \\ Delft University of Technology \\ Delft, The Netherlands\\
s.h.tindemans@tudelft.nl}\vspace{-3em}
}

\IEEEpubid{
\begin{minipage}{0.5\textwidth}
\begin{flushleft}
\vspace{1cm}
978-1-5386-3596-4/18/\$31.00~\copyright~2018 IEEE 
\end{flushleft}
\end{minipage}
\begin{minipage}{0.5\textwidth}
\begin{flushright}
\vspace{1cm}
    PMAPS 2018
\end{flushright}
\end{minipage}}

\maketitle

\begin{abstract}

Machine learning techniques have been used in the past using Monte Carlo samples to construct predictors of the dynamic stability of power systems. In this paper we move beyond the task of prediction and propose a comprehensive approach to use predictors, such as Decision Trees (DT), within a standard optimization framework for pre- and post-fault control purposes. In particular, we present a generalizable method for embedding rules derived from DTs in an operation decision-making model. We begin by pointing out the specific challenges entailed when moving from a prediction to a control framework. We proceed with introducing the solution strategy based on generalized disjunctive programming (GDP) as well as a two-step search method for identifying optimal hyper-—parameters for balancing cost and control accuracy. We showcase how the proposed approach constructs security proxies that cover multiple contingencies while facing high-dimensional uncertainty with respect to operating conditions with the use of a case study on the IEEE 39-bus system. The method is shown to achieve efficient system control at a marginal increase in system price compared to an oracle model.
\end{abstract}


\section{Introduction}\label{sec:intro}
\vspace{-0.5em}
The increasing uncertainty that surrounds system operation renders the adoption of probabilistic security assessment frameworks a high priority for many Transmission System Operators (TSOs) worldwide. In the past, large-scale Monte Carlo techniques that involve the high-density sampling of operating points and post-fault stability assessment via time-domain simulations have been proposed (e.g., \cite{Weh98,Kri11,Kon16}). Consequently, machine learning can be applied to the Monte Carlo samples to build rules that predict post-fault stability for unseen operating points. For this predictive task, most researchers have adopted decision trees (DT) or DT ensembles.

Beyond prediction of post-fault stability, such sample-derived rules can also be used as a control method to delineate the system's pre-fault stable operating region. By embedding appropriate constraints in a TSO's operation and scheduling tools, post-fault stability can, in theory, be achieved. In general, two different approaches can be used to infer suitable control actions from sample-derived rules; a heuristic search strategy and an optimization-based approach. 
In the heuristic followed by \cite{Kar02,Xu14}, when a post-fault unstable operating point is encountered, the DT's decision path of generator power bounds is followed upwards from the terminal node to the corresponding parent node. Subsequently, the generators are redispatched according to the threshold of this parent node to shift the operation to the child node encapsulating mostly post-fault stable operating points.
The second approach, followed by e.g., \cite{Gen10,Cos16}, solves an Optimal Power Flow (OPF) problem for each terminal node encapsulating mostly post-fault stable points, where the corresponding decision path has been added in the form of inequality constraints. After solving all constrained OPFs, the solution with minimal operating cost is selected. Recently, the authors of \cite{Tha17} published a Security-Constrained-OPF (SCOPF) based on data considering line flow limits as features. A slightly different approach, but one that still considers the complete decision path, is proposed in \cite{Liu14}. Post-fault stable generation re-dispatch is achieved by first finding the most effective generators and second restricting the assumed post-fault stable region with adapted bounds for the generator powers. 

In practice two challenges arise: (i) the online computation of current control approaches using rules from Monte Carlo samples entails a high computational burden, (ii) the very nature of cost optimality drives system operation right on the limiting rule \cite{Gen10}, thus potentially leading to post-fault unstable operation even in cases of DTs with very high prediction accuracy.

Apart from describing the challenges of using sample-derived rules in a control setting, the contributions of this paper are twofold:
\begin{itemize}
\item Introducing the disjunctive formulation of the control approach to reduce computational complexity
\item Proposing a procedure to select parameters that improve the accuracy of rules in a control setting. 
\end{itemize}

A novel approach to embedding disjunctive rules for feasible operation is introduced to address the issue of computational complexity. In particular, the standard corrective operation problem can be extended using generalized disjunctive programming (GDP) \cite{Gro94}, resulting in a mixed integer linear problem (MILP) for the DC case or a mixed integer nonlinear problem (MINLP) in the AC case. Convex-hull and Big-M reformulations are considered to account for the disjunction of learned convex regions. The computational benefit of the proposed approach results from the GDP formulation that enables solvers to make use of branch--and--bound search to efficiently identify the global optimum with respect to the implemented rules instead of evaluating the disjunctive convex regions one by one.

Moreover, a two-step parameter search is proposed to address the challenge of rule inaccuracy: k-fold cross validation is used to balance under and overfitting and a safety margin is computed to finally ensure rule accuracy for the complete uncertainty spectrum.

The specific challenges and the elements of the approach are illustrated using an IEEE 39 bus case study. In this paper, a steady state DC analysis is used instead of time domain AC simulations. 
This allows to compare the proposed approach against a globally cost-optimal reference point (obtained by the SCOPF solution) thus simplifying the reproducibility. The case study is used to compare both, the computational benefits and the cost-effectiveness.

The rest of the paper is structured as follows. In Section \ref{sec:problem}, the objective of the control purpose and two specific challenges are illustrated. Subsequently, in Section \ref{sec:control}, the disjunctive approach is introduced including the parameter search for correcting the inaccuracy of disjunctive rules. 
The case study is carried out in Section \ref{sec:results} and Section \ref{sec:conc} concludes the paper.
\section{Embedding Rules in Control}\label{sec:problem}
\vspace{-0.5em}
\subsection{Objective}\label{subsec:obj}
\vspace{-0.5em}
In broad terms, the objective is to build a control approach to finding an acceptable and cost-optimal operating point of the power system given the training data $(X,Y)$ containing $n$ samples $(x_i,y_i)$, $i=1,\dots n$ of operating points, each with $p$ features $x_i \in \mathbb{R}^p$ and a class label $y_i \in \{0,1\}$, where $y_i=1$ and $y_i=0$ corresponds to acceptable and unacceptable operating points, respectively. $X$ is assumed to be generated randomly using a Monte Carlo sampling process and is representative for expected operating conditions. The binary class label acceptable/unacceptable can be obtained from simulations and correspond to a user-specified stability criterion. 

The objective of this paper is to find rules from $(X,Y)$ in the form of inequality constraints $k_{xy}(x,z)\leq 0$ limiting a control approach to acceptable operations. These rules will be embedded in the OPF optimization
\vspace{-0.5em}
\begin{equation}\label{eq:ObjectiveOPF}
\begin{aligned}
& \underset{x,z}{\text{min}}
&& f(x) && \\
& \text{s.t.}
&& \tilde{h}(x) = 0  \\
& && \tilde{g}(x) \leq 0 \\
& && k_{xy}(x,z) \leq 0,
\end{aligned} \vspace{-0.5em} 
\end{equation}
where $x$ contain all power system variables, such as generator powers, line flows, phase angles and bus voltages and $z$ are any other auxiliary variables. $f(x)$ comprise the system's operation cost, $h(x)$ are the typical equality constraints, such as node balances, line flow equations and $g(x)$ are the inequality constraints for physical limitations of the system. Note, the tilde symbol $\sim$ is denoted to the operational uncertainty surrounding the system (e.g., load injections can vary from sample to sample). In the ACOPF case, variables and constraints for active and reactive power are considered, whereas in the DCOPF case only constraints involving active power must be considered. We would like to emphasize: it is not the objective to provide a full control approach involving real post-fault stability assessments and an AC setting, rather to obtain $k_{xy}(x,z)$ from $(X,Y)$ in an efficient and accurate way with respect to the purpose of controlling the system. Through this purpose and through the fact of straddling both domains, mathematical optimization and machine learning, specific challenges arise. 

\vspace{-0.5em}
\subsection{Challenges}\label{sec:chall}
\vspace{-0.5em}
Current methods to learn Monte Carlo sample-derived rules for control purposes face two particular challenges, which do not arise when using rules solely for prediction: the online computational complexity and the inaccuracy of rules.

\begin{figure}
\centering
   \begin{subfigure}[b]{0.3\textwidth}
   \includegraphics[width=1\linewidth]{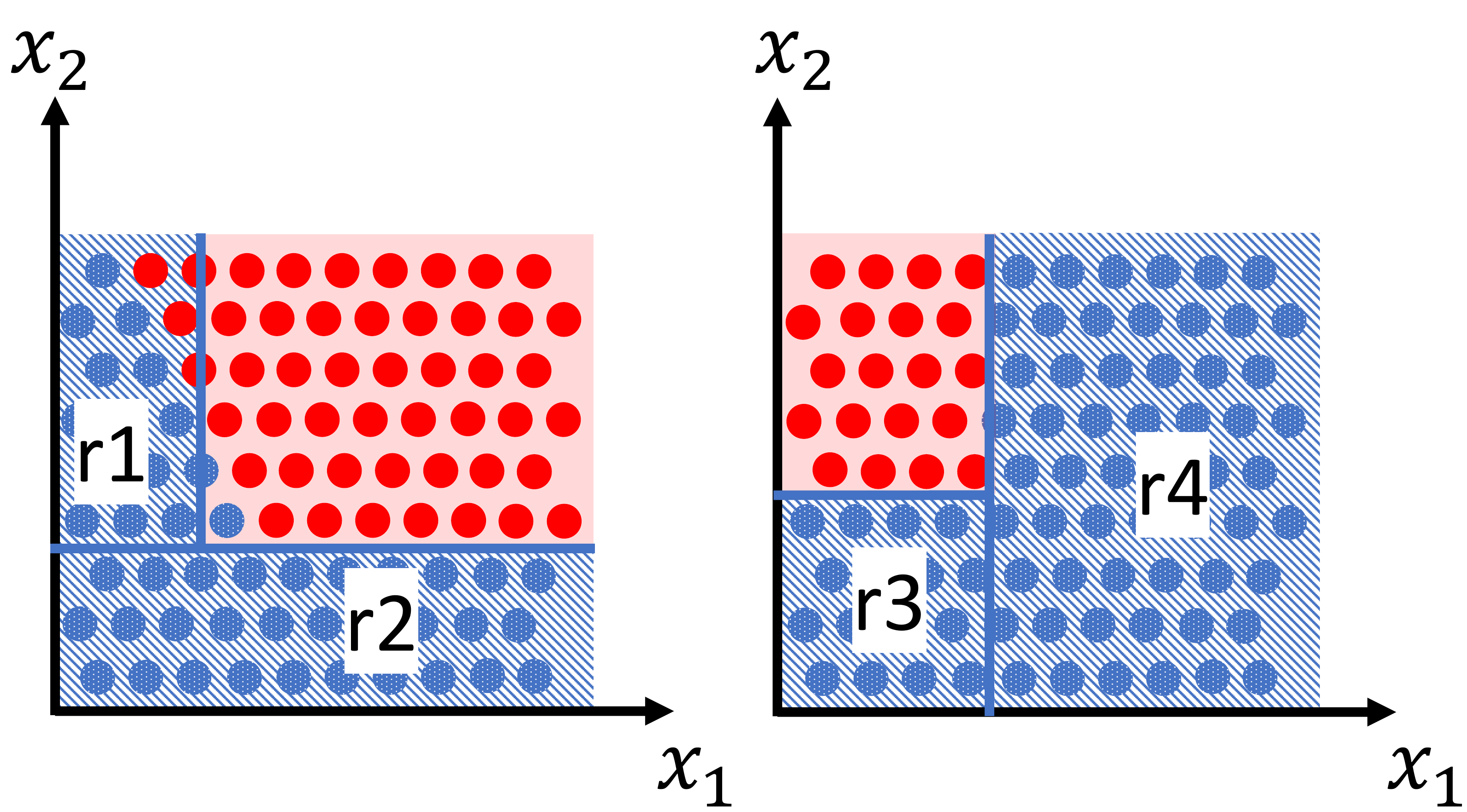}
   \caption{}
   \label{fig:DTsingle} 
\end{subfigure}
\begin{subfigure}[b]{0.15\textwidth}
   \includegraphics[width=1\linewidth]{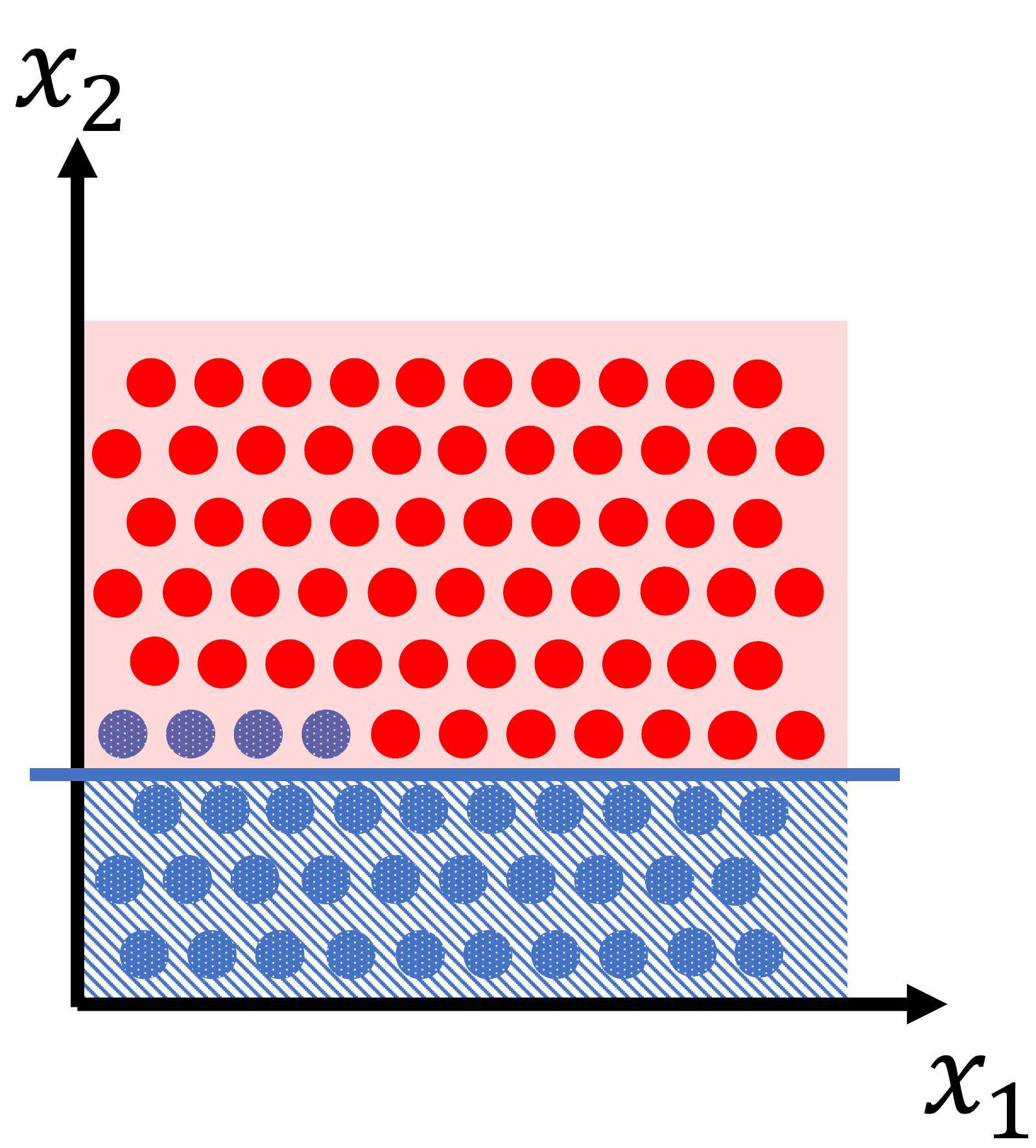}
   \caption{}
   \label{fig:DTglobal}
\end{subfigure}
\caption{\footnotesize Learning DT rules for two faults using (a) two single DTs (one per fault) resulting in total in four rules (blue lines) and (b) one global DT resulting in one rule. Acceptable and unacceptable operating points are indicated in blue and red.}\vspace{-1.5em}
\end{figure}

\subsubsection{Computational complexity}
Current prediction-oriented approaches 
typically comprise an offline training part, where a group of classifiers is trained on the data and an online part, where the current operating point of interest is evaluated using the pre-trained classifiers. Usually, one DT is trained per fault since each fault can have specific characteristics (e.g. individual critical features and individual nonlinear boundary). 
The complexity to evaluate one operating point in one fault-specific tree is $\mathcal{O}(n)$ where $n$ is the maximal tree depth; consequently, to evaluate against $c$ potential faults involves computations with $\mathcal{O}(c \, n)$.  
This computational overhead of evaluating an operating point is negligible since it consists of evaluating simple algebraic equations. 

In the online part (as in e.g. \cite{Gen10,Cos16}), the computations are more costly since many optimizations (one per terminal node $t$) have to be solved and achieving close to real-time performance can pose substantial challenges. 
In particular, when learning multiple trees (one tree per fault $c$) and combining the rules across all trees (e.g. as in \cite{Xu14}) results in high computational complexity with an exponential growth $\mathcal{O}(t^c)$ in the number optimizations to be solved.
E.g, the example in Fig. \ref{fig:DTsingle} involves two unacceptable faults in terms of pre-fault operation state variables $x_1$ and $x_2$ and the final acceptable feasible operating region (shaded blue area) is the conjunction of the areas \{(r1,r3),(r1,r4),(r2,r3),(r2,r4)\}; this requires to solve  four optimization problems, one for each conjunction. 
In contrast, the computational complexity gets reduced to $\mathcal{O}(t)$ optimizations when using one global tree for overall acceptability. In the example, the four linear rules from Fig. \ref{fig:DTsingle} are reduced to a single linear rule for the blue shaded area in Fig. \ref{fig:DTglobal} and resulting in solving a single optimization. 
A single global DT is also used in \cite{Weh98} and \cite{Cos16,Liu14}. It resulted in a rough reduction of $\SI{80}{\percent}$ in total number of terminal nodes to be considered in \cite[p.~205]{Weh98}. 
Despite the computational reduction to solve $\mathcal{O}(t)$ optimizations, this still represents a significant computational burden for the online workflow. 
\begin{figure}
\centering
\includegraphics[scale=.19]{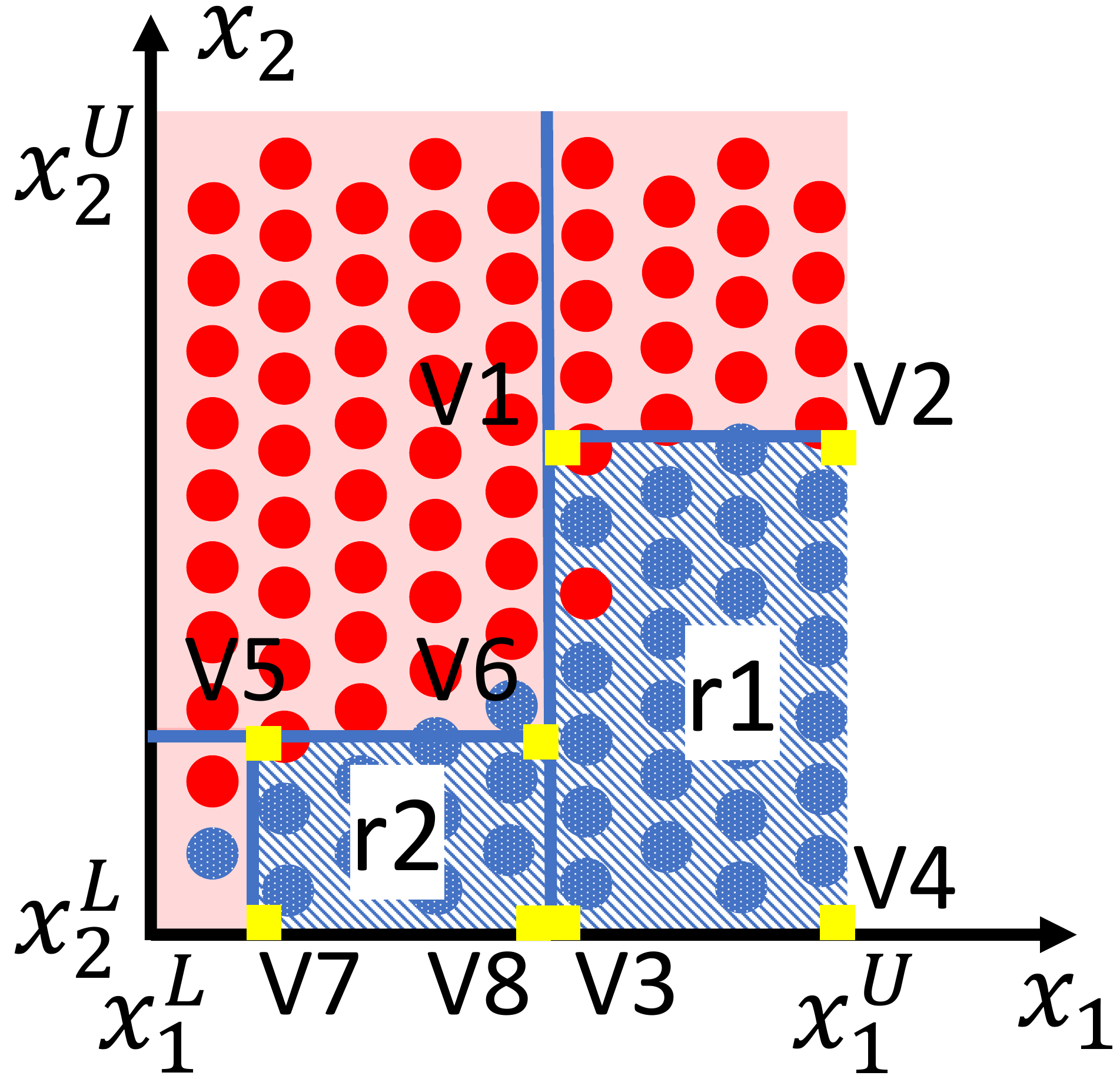}
\caption{\footnotesize Candidate points at the vertices (shown in yellow) of the inequality constraints.}
\label{fig:optimalonedges}
\vspace{-1.5em}
\end{figure}

\subsubsection{Accuracy of rules for control}\label{sec:acc}
The goal of using rules for control is to minimize the operating points obtained by the control model and falsely classified as acceptable - let's call this control error. 
Without unnecessarily restricting the operating region
When learning a tree, the algorithm aims to minimize the test error for the given training population $(X,Y)$ by using e.g. the Gini impurity. However, it is very unlikely that this test error equals the control error (as assumed e.g. in \cite{Kar02}) since both metrics refer to radically different populations.
In fact, the operating points in the control problem are the result of the optimization stated in Section \ref{subsec:obj} where the rules are implemented as linear inequality constraints $k_{xy}(x,z) \leq 0$. 
As such, these operating points are locally concentrated in specific areas of the state-space since the optimal point in a convex linear problem always lies on one of the vertices of the feasible region or on the line segment between two vertices if both are optimal (fundamental theorem of linear programming).
For instance, let us consider Fig. \ref{fig:optimalonedges}. 
Even if the test error is low, e.g. for region r1 (enveloped by vertices V1-V4) the test error is $\SI{7.7}{\percent}$ (since $2$ out of $26$ points are wrongly labelled by the rules as acceptable), the local error for vertex V1 is at $\SI{100}{\percent}$. 

To deal with this challenge, \cite{Kar02} proposed to heuristically determine a new unit commitment without cost considerations after the re-dispatched point is assessed as unacceptable.
\cite{Tha17,Liu14} proposed to address this issue in the DT learning process by asymmetrically adjusting the weights of observations' prior probabilities resulting in a conservative shift of the rules towards the acceptable region; this might lead to over or underfitting the data (in particular DTs are known to need tuning to avoid overfitting \cite{Jam14}).
In fact, large factors (e.g., $0.99$ and $0.01$ in \cite{Liu14} for unacceptable and acceptable classes respectively) were used and no methodology for deriving this numbers has been presented. As in \cite{Xu14}, the challenge of rule inaccuracy can also be addressed in the online work-flow by incrementally scanning along a margin to the offline-learned rule until an acceptable operating point is identified. 
This online scan requires evaluating the label of the operating points for all faults 
and would result in a prohibitively high computational burden for online control applications in the case of considering post-fault stability as acceptability criterion.

\section{Disjunctive approach for control}\label{sec:control}
\begin{figure}
\centering
   \begin{subfigure}[b]{0.26\textwidth}
   \includegraphics[width=1\linewidth]{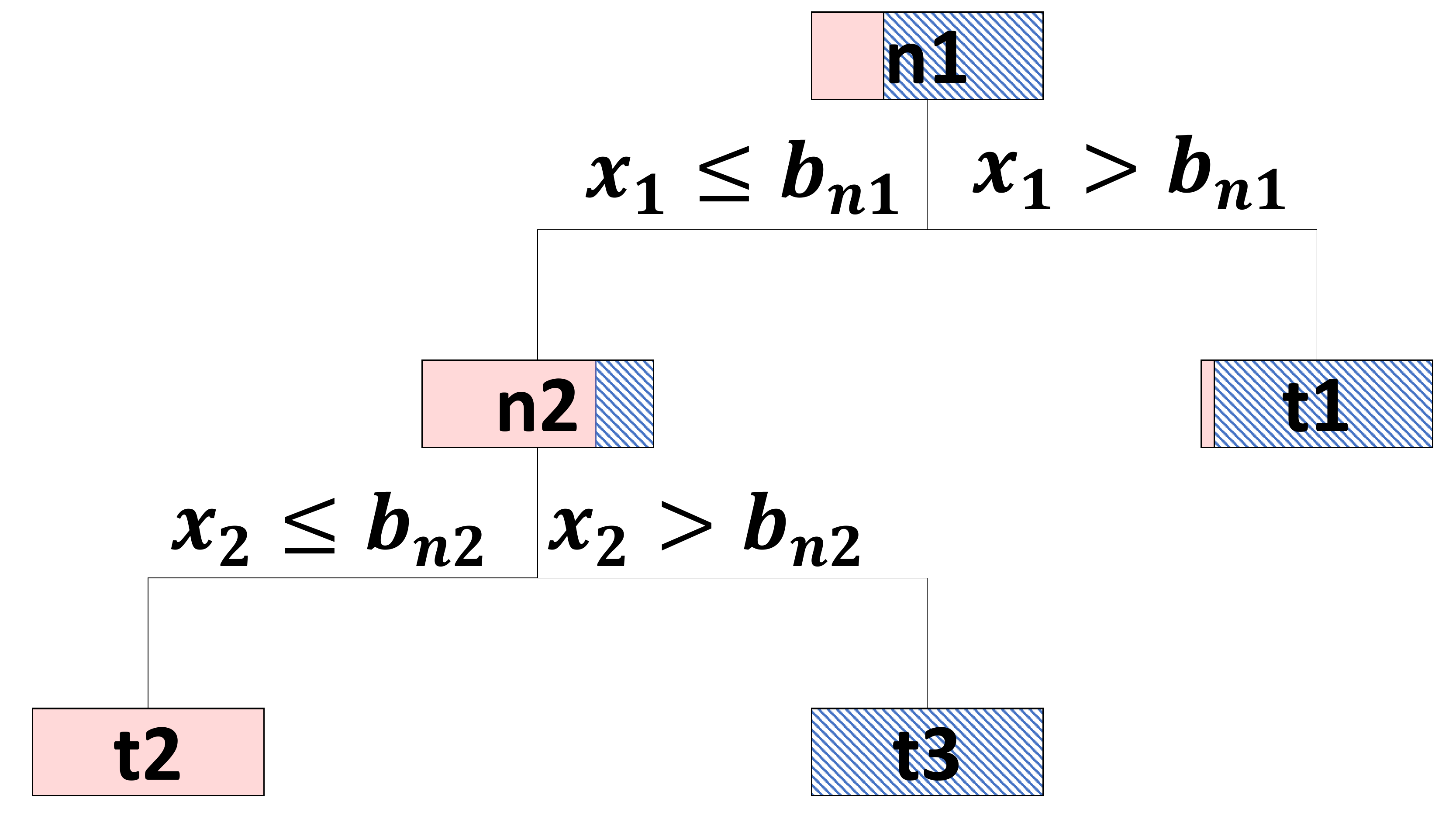}
   \caption{}
   \label{fig:DTschematic1} 
\end{subfigure}
\begin{subfigure}[b]{0.26\textwidth}
   \includegraphics[width=1\linewidth]{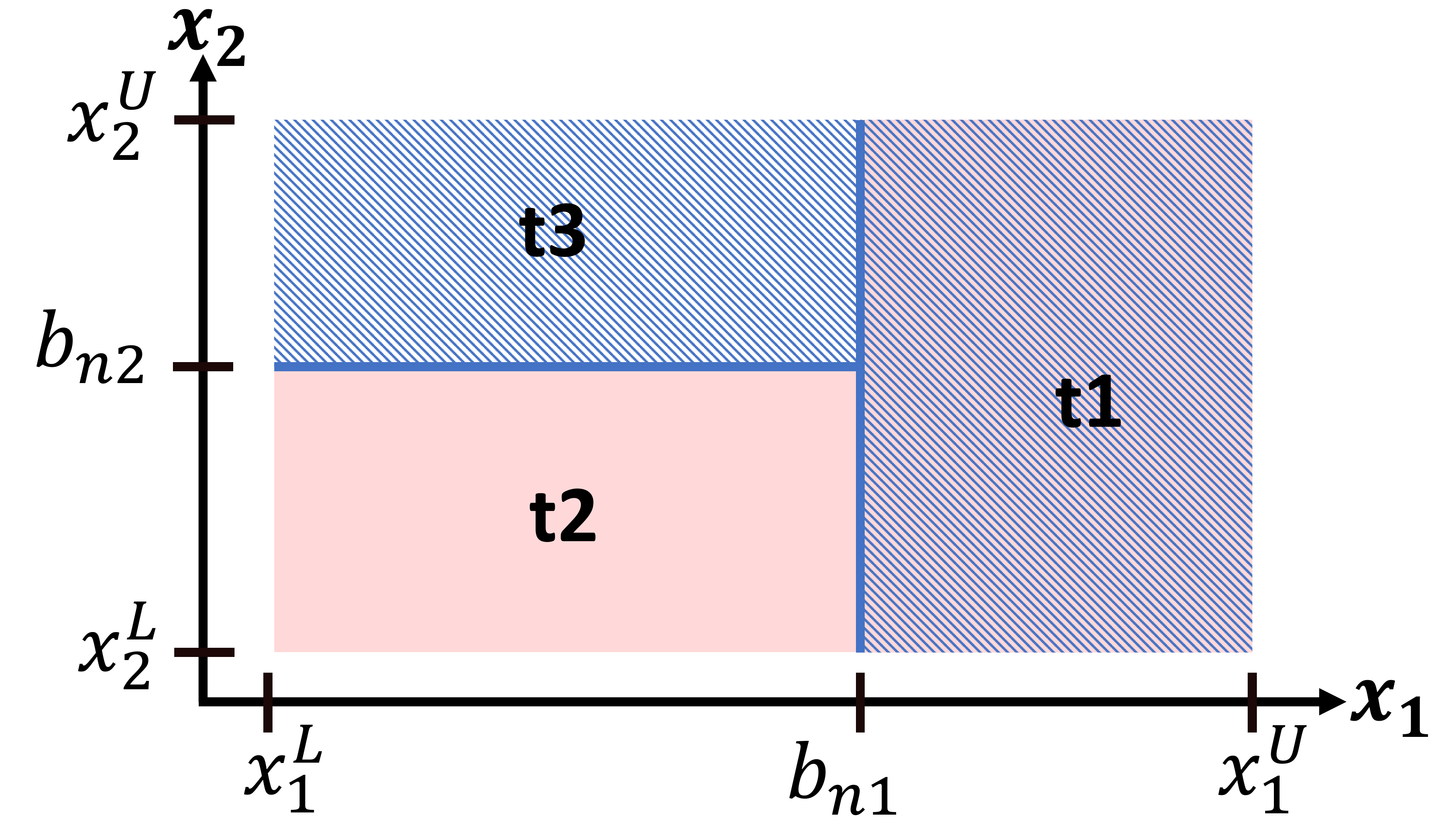}
   \caption{}
   \label{fig:DTschematic2}
\end{subfigure}
\caption{\footnotesize (a) Schematic tree structure and (b) corresponding splits in the feature space as blue lines. Acceptable and unacceptable regions/nodes are indicated in dashed blue and light red, respectively.}\label{fig:DTschematic}\vspace{-1.5em}
\end{figure}

\vspace{-0.5em}
\subsection{Obtaining rules using machine learning}\label{subsec:conobt}
\vspace{-0.5em}
For binary classification, the CART learning algorithm \cite{Bre84} successively splits the feature space in two half-spaces in each iteration based on the training data $(X,Y)$. Considering a univariate DT, let $a_n \in \{0,1\}^p$ be the single-entry vector of each branch node $n$ corresponding to the split position in the feature vector $x \in \mathbb{R}^p$ and $b_n \in \mathbb{R}$ be the split threshold of $n$. The example of Fig. \ref{fig:DTschematic} illustrates the splitting scheme using $p=2$. The algorithm starts by finding the first best split $a_{\mathrm{n1}}^{\intercal} = (1,0)$ for the initial branch node n1 with threshold $b_{\mathrm{n1}}$. Consequently, the regions $a_{\mathrm{n1}}^{\intercal} x \leq b_{\mathrm{n1}}$ and $a_{\mathrm{n1}}^{\intercal} x > b_{\mathrm{n1}}$ contain the unacceptable and acceptable classes, respectively. 
The algorithm terminates in the region $a_{\mathrm{n1}}^{\intercal} x > b_{\mathrm{n1}}$ since the purity criterion (e.g. gini impurity is less than a user-defined threshold) is met leading to a terminal node t1.
In contrast, $a_{\mathrm{n1}}^{\intercal} x \leq b_{\mathrm{n1}}$ is not pure enough thus the algorithm continues in this branch to split in the same way until a stopping criterion holds (e.g., limit of tree depth, pureness of nodes, etc.). In total, this example tree has two branch nodes n1, n2, two acceptable terminal nodes t1, t3 and one unacceptable terminal node n2 corresponding to the three regions in Fig. \ref{fig:DTschematic2}. 

\begin{figure}
\centering
   \begin{subfigure}[b]{0.39\textwidth}
   \includegraphics[width=1\linewidth]{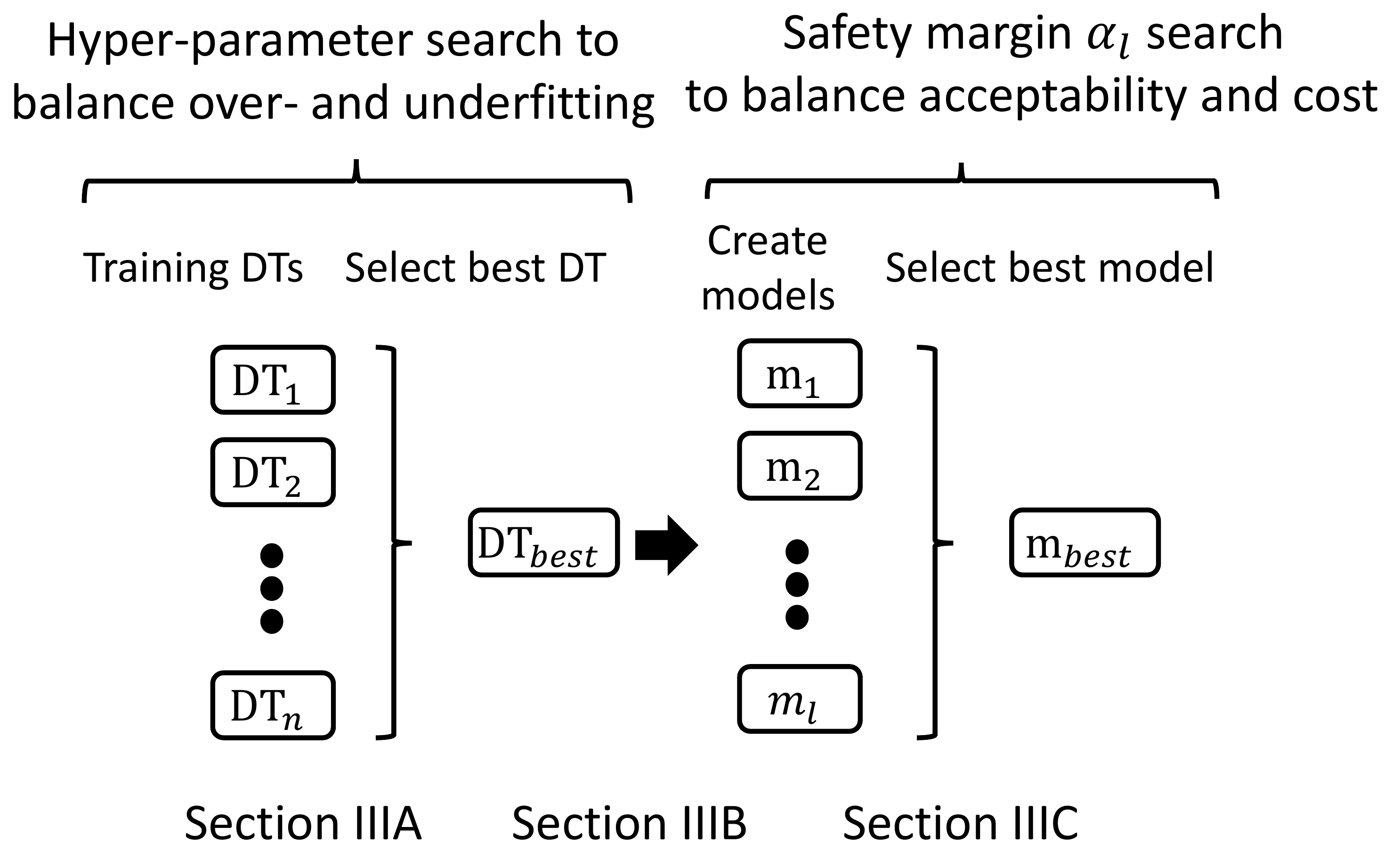}
   \caption{}
   \label{fig:procedureoffline} 
\end{subfigure}
\begin{subfigure}[b]{0.39\textwidth}
   \includegraphics[width=1\linewidth]{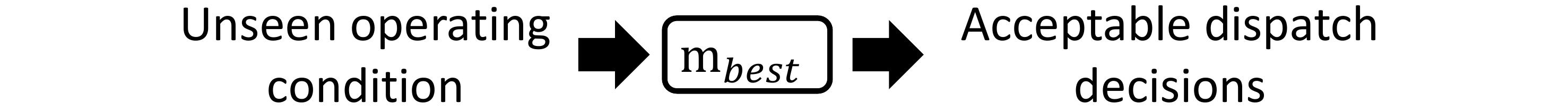}
   \caption{}
   \label{fig:procedureonline}
\end{subfigure}
\caption{\footnotesize Procedure of (a) the offline and (b) the online part.}\label{fig:procedure}\vspace{-1.5em}
\end{figure}

To balance under- and overfitting, a hyper--parameter grid-search is proposed by using an n-fold cross-validation learning procedure (Fig. \ref{fig:procedureoffline}). Many DTs ($\mathrm{DT}_1 \dots \mathrm{DT}_n$) are trained through CART for many hyper-parameter combinations using the given training population $(X,Y)$. These may include the maximal tree depth, the minimal number of samples in one terminal node, the terminal node's Gini impurity and/or a maximal number of terminal nodes. The best performing $\mathrm{DT}_{best}$ is selected through out-of-sample testing by the use of a user-specified criterion (e.g, typically the \textit{'f1'} score or alternatively the classical test error).

\vspace{-0.5em}
\subsection{Accounting for disjunctive rules in the optimization}
\vspace{-0.5em}
To reduce the online computations, the necessity of solving multiple optimizations is avoided (e.g., \cite{Gen10,Cos16} solves one optimization for each terminal node encapsulating the acceptable class).
Here, the rules are accounted in a single optimization as a disjunction. The mutually exclusive disjunction contains one rule for each terminal node $t \in T_A$ of the acceptable class. Each rule corresponds to all inequality constraints of all branch nodes from the initial node to the terminal node.
Consequently, in the example in Fig. \ref{fig:DTschematic}, two optimizations that take into account either the rule $a_{\mathrm{n1}}^{\intercal} x > b_{\mathrm{n1}}$ for t1 or the rule $a_{\mathrm{n1}}^{\intercal} x \leq b_{\mathrm{n1}} \, \land \, a_{\mathrm{n2}}^{\intercal} x > b_{\mathrm{n2}}$ for t3 is replaced by one optimization accounting for the disjunction $(a_{\mathrm{n1}}^{\intercal} x > b_{\mathrm{n1}}) \lor (a_{\mathrm{n1}}^{\intercal} x \leq b_{\mathrm{n1}} \, \land \, a_{\mathrm{n2}}^{\intercal} x > b_{\mathrm{n2}})$.

To reformulate disjunctions for optimizations, two different approaches from GDP can be adopted involving binary variables. Whereas the so-called Big-M reformulation results in fewer variables and constraints, the convex-hull reformulation \cite{Bal85} results in a relaxed linear problem with a feasible region at least as tight as the one from Big-M reformulation \cite{Gro94}. Since in this univariate DT case the disjunctions are all simple and all variables are bounded, the Big-M formulation results in the same tight relaxation. Let us define $A_L(t)$ as the set of ancestor branch nodes whose left branch has been followed on the path from initial node to the terminal node $t$. Similarly, $A_R(t)$ is the set of right-branch ancestors. The sets of the example of Fig. \ref{fig:DTschematic} would be $A_R(\mathrm{t1})=\{\mathrm{n1}\}$, $A_L(\mathrm{t1})=\{\}$, $A_R(\mathrm{t3})=\{\mathrm{n2}\}$ and $A_L(\mathrm{t3})=\{\mathrm{n1}\}$.
The reformulation is presented below:
\vspace{-0.5em}
\begin{subequations}\label{eq:bigm}
\begin{align}
& a_{n}^{\intercal} x \leq b_n z_t + a_{n}^{\intercal} M_1 (1-z_t) && \forall t \in T_A, \, \forall n \in A_L(t) \label{eq:inequality}\\
& a_{n}^{\intercal} x > b_n z_t + a_{n}^{\intercal} M_2 (1-z_t), && \forall t \in T_A, \, \forall n \in A_R(t) \label{eq:strictinequality}
\end{align}
\end{subequations}
where $z_t \in \{0,1\}$ is a binary optimization variable for all terminal nodes  encapsulating the acceptable class $\forall t \in T_A$. Further, the optimization is enforced to assign exactly one accepted terminal node by
\vspace{-0.5em}
\begin{equation}\label{eq:terminalnode}
\begin{aligned}
\mathop{\mathlarger{\sum}} \limits _{t \in T_A} z_t = 1.
\end{aligned} \vspace{-0.5em}
\end{equation}
Since the use of a strict inequality in Equation \eqref{eq:strictinequality} is not possible in mixed integer optimizations, we propose to add a small value $\gamma \in \mathbb{R}_{>0}$ to change to a non-strict inequality:
\vspace{-0.5em}
\begin{equation}
\label{eq:nonstrictinequality}
a_{n}^{\intercal} x \geq b_n z_t + a_{n}^{\intercal} M_2 (1-z_t) + \gamma \quad \forall t \in T_A, \, \forall n \in A_R(t). \vspace{-0.5em}
\end{equation}
Note, $\gamma\in \mathbb{R}_{>0}$ should be selected in accordance with the solver sensitivities since selecting it too small could cause numerical instabilities in the solver. The vectors of big-$M$ constants $M_1 \in \mathbb{R}^p$ and $M_2 \in \mathbb{R}^p$ have to be selected smallest as possible that the relaxed problem has a small feasible region in order to speed-up computations. The smallest possible big-$M$ constants are obtained by
\vspace{-0.5em}
\begin{subequations}\label{eq:bigmselection}
\begin{align}
& M_1 = \mathrm{max} \{ a_n b_n + \overline{a}_n x^L : n \underset{t \in T_A}{\cup} A_L(t) \} \\
& M_2 = \mathrm{min} \{ a_n b_n + \overline{a}_n x^U: n \underset{t \in T_A}{\cup} A_R(t) \},
\end{align}
\end{subequations}
where $\overline{a}_n=1-a_n$ is the negation of $a_n$,
$\mathrm{max}$ and $\mathrm{min}$ are element-wise comparison operators and all considered variables were assumed to be bounded $x^L \leq x \leq x^U$.

The computational benefit of this formulation results from the following: (i) only one model must be initialized, (ii) only one pre-solve step is required, (iii) the solver can make use of branch--and--bound search, (iv) the solver still can further accommodate speed-up methods (e.g. decomposition or batch methods). In addition, reduction in computation is achieved by learning a single tree for global acceptability instead of learning multiple fault--specific trees to reduce the total number of terminal nodes.

\vspace{-0.5em}
\subsection{Correction of control-oriented rules}\label{sec:learn}
\vspace{-0.5em}
\begin{figure}
\centering
\includegraphics[scale=.19]{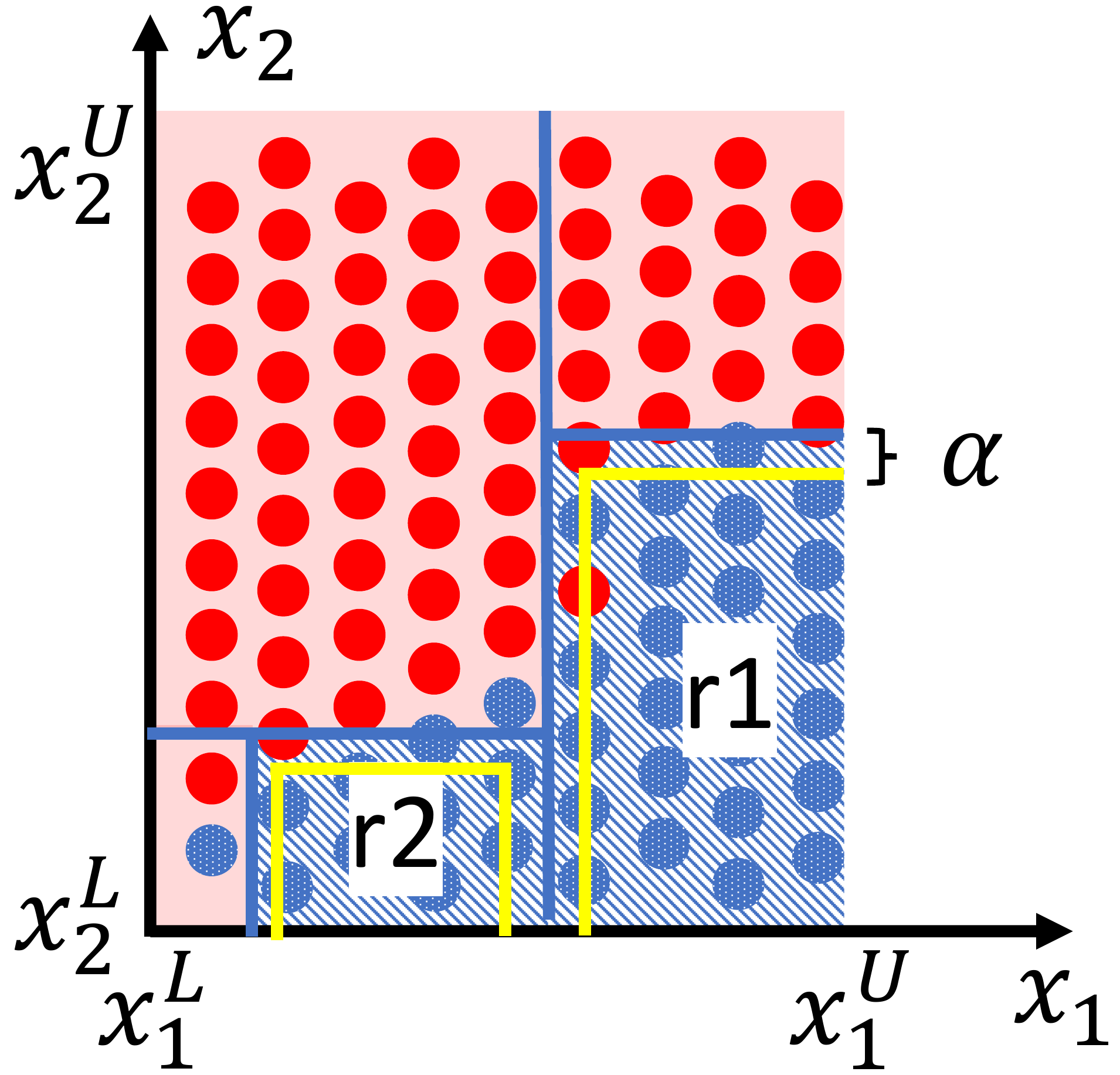}
\caption{\footnotesize Rules for control purpose: original DT rules as blue lines and rules with safety margin $\alpha_l$ as yellow lines.}
\label{fig:alphaapproach}\vspace{-1.5em}
\end{figure}

To address the issue of rule accuracy in a control setting (as discussed in Section \ref{sec:acc}), a safety margin $\alpha_l$ is used to shift the rule towards the acceptable region as illustrated in Fig. \ref{fig:alphaapproach}. For increasing $\alpha_l$, the two regions r1 and r2 of the figure are narrowed and therefore the adapted rules provide more reliability since the edges lie further inside the acceptable region. To account for the safety margins, the two inequality constraints Equations \eqref{eq:inequality}, \eqref{eq:nonstrictinequality} are adapted to
\vspace{-0.5em}
\begin{subequations}\label{eq:bigm3}
\begin{align}
 a_{n}^{\intercal} x \leq (b_n - \alpha_l) z_t + a_{n}^{\intercal} M_1 (1-& z_t) \nonumber\\ 
 & \forall t \in T_A, \, \forall n \in A_L(t) \\
 a_{n}^{\intercal} x \geq (b_n + \alpha_l) z_t + a_{n}^{\intercal} M_2 (1- & z_t) + \gamma. \nonumber \\
& \forall t \in T_A \, \forall n \in A_R(t)
\end{align}
\end{subequations}
The safety margin $\alpha_l>0$ that guarantees acceptable operation for the complete range of uncertainty is searched offline (Fig. \ref{fig:procedureoffline}). Hence, the critical computations of evaluating the acceptability label of an operating point are shifted from online (e.g., as in \cite{Xu14}) to offline. For this parameter search, $L$ different constrained optimization models $\mathrm{m}_l$ with varying $\alpha_l \, \forall l \in L $ are initialized and each $\mathrm{m}_l$ is assessed by the use of a set of samples drawn from the distribution of uncertain operation.
For those samples, the dispatch decisions are computed by solving all created $\mathrm{m}_l$. For each of the dispatch decisions (samples $\times$ $\mathrm{m}_l$), the acceptability label is assessed. 
Finally, the best $\mathrm{m}_{best}$ with the respective $\alpha_l$ is selected by the TSO in consideration of the minimization of total operating cost (e.g., lost load and generation cost). Note, $\mathrm{m}_{best}$ is the only model used in the online computations and the scale of the search for $\mathrm{m}_{best}$ (and $\alpha_{best}$) depends on the user's experience of the specific accuracy of the learning algorithm. In this paper, an exhaustive search is used for the purpose of studying the effects of $\alpha_l$ in the case study.

\section{Case study}\label{sec:results}

\vspace{-0.5em}
\subsection{Test system and solution strategy}
\vspace{-0.5em}
\begin{table}
\begin{center}
\centering 
\caption {\footnotesize Generator data of the IEEE 39 bus system.}\label{tab:generators}
\resizebox{.49\textwidth}{!}{
\begin{tabular}{c|cccccccccc}
\toprule
\footnotesize
$g$	& $\mathrm{G0}$ & $\mathrm{G1}$ & $\mathrm{G2}$ & $\mathrm{G3}$ & $\mathrm{G4}$ & $\mathrm{G5}$ &  $\mathrm{G6}$ & $\mathrm{G7}$ & $\mathrm{G8}$ & $\mathrm{G9}$ \\
\midrule
$p^{U}_{g}$ [$\si{\MW}$] & 1040 & 725 & 652 & 508 & 687 & 580 & 564 & 865 & 1100 & 4000 \\
$\epsilon_g$ [$\si{\dollar\per\MWh}$] & 30 &24 &26 &32 &34 &42 &44 &46 &48 &35 \\
\bottomrule
\end{tabular}}
 \vspace{-2em}
\end{center}
\end{table}  

\begin{figure}
\centering
\includegraphics[scale=0.25]{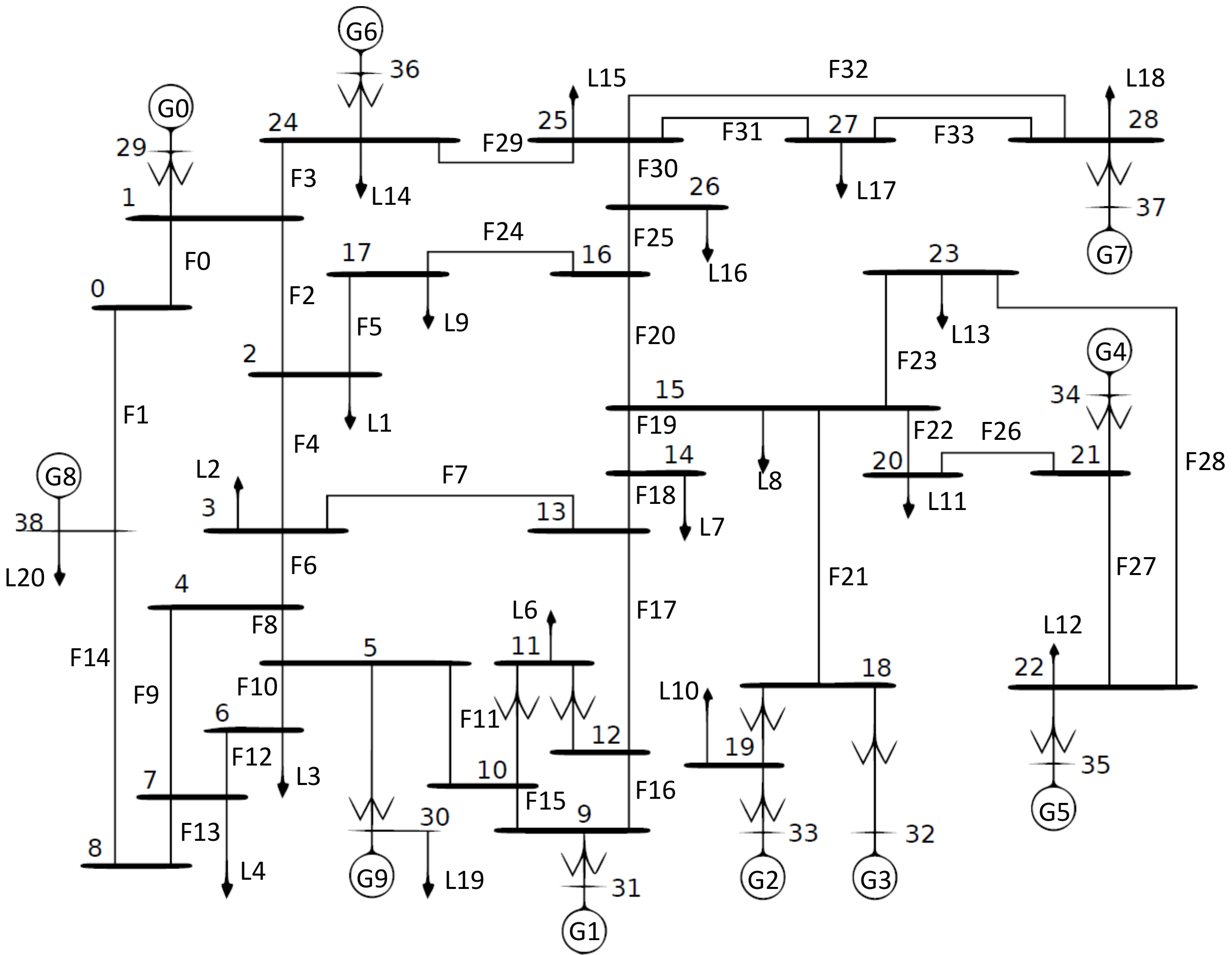}
\caption{\footnotesize IEEE 39 bus New England system.}
\label{fig:IEEE39BusNetwork}
\vspace{-1.5em}
\end{figure}

The objective of this study is to demonstrate the methodology presented in this paper. The focus laid on the relevant aspects corresponding to the challenges caused by striving both domains, machine learning and mathematical optimization. Consequently, a DC power flow approximation were used and an operating point was considered to be acceptable if no post-fault loss of load occurs following any of the possible faults, while the list of faults analyzed included all $N-1$ line outages (similar assumptions have been made by several authors e.g. \cite{Gen10}). These assumptions allowed a comparison with a reference solution (a global cost minimum among acceptable points) obtained using a direct solution of the SCOPF problem.
 
The IEEE 39 bus system was used in this case study. 
The network connectivity is shown in Fig. \ref{fig:IEEE39BusNetwork} and data such as nominal loads and line reactances were taken from \cite{Ana12}. The system was modified in some aspects: to ensure the feasibility of N-1 secure solutions for the complete uncertainty spectrum, all generators allowed for post-fault corrective redispatches of $\pm \SI{100}{\MW}$ and all line flow limits were set to $\SI{2000}{\MW}$. 

To generate a training set, operational uncertainty was considered in all loads with a deviation of $\pm \SI{25}{\percent}$ from the nominal loads. Load samples were generated from a multivariate Gaussian distribution with Pearson's correlation coefficient of $0.75$ between each load pair. The inverse transformation method was used to convert sampled values to a marginal Kumaraswamy distribution with the probability density function
\begin{equation}\label{eq:pdf}
\begin{aligned}
f(x)=abx^{a-1}\left(1-x^a\right)^{b-1},\nonumber
\end{aligned}
\end{equation}
where $a=1.6$, $b=2.8$ and $x \in [0,1]$. Finally, the sampled values were scaled to the desired range of load variation. The generator power levels were randomly sampled from an uncorrelated uniform distribution within their specific operating range (lower bound was $p^{L}_{g}=0$ for all generators $g$ and upper bounds $p^{U}_{g}$ are shown in Table \ref{tab:generators}). Since no power losses were assumed, the total power of loads must equal the total power generation. Any mismatch was distributed over all generators (positive as negative) by weighting based on the capacity of the generators $p^{U}_{g}$. All samples that led to a physical inconsistency were disregarded. 

In the hyper--parameter grid-search under- and overfitting were balanced for the DT (Fig. \ref{fig:procedure}) via 5-fold cross validation. $500000$ samples were used in the training set and the $65$ features consisted of all generator, loads and line flow power levels. Each indivual DT for the global acceptability was learned using the CART algorithm \cite{Bre84} (as implemented in the package \textit{scikit-klearn} 0.18.1 \cite{Ped11} with Python 3.5.2). The best split was selected successively based on minimizing the Gini impurity and the selected non--default settings were to balance different population sizes in two classes and the parameters involved in the hyper--parameter grid-search: the maximal DT depth $[5,6, \cdots 20]$ and the maximal number of terminal nodes $[20,40,\cdots 100,200,\cdots 500]$. The best $\mathrm{DT}_{best}$ was selected using the \textit{'f1'} score; the optimal parameters found were a maximal tree depth of $14$ and a maximal number of terminal nodes of $200$. The final resulting classifier had $82$ acceptable and $118$ unacceptable terminal nodes. 

The safety margin $\alpha_l$ was searched to balance acceptability and cost (Fig. \ref{fig:procedure}). Here, an exhaustive search was undertaken by varying the safety margin in $\alpha_l \in [0,\SI{60}{\MW}]$ with a step-size of $\SI{1}{\MW}$ to study its effect and resulted in $L=61$ models $\mathrm{model}_l$ for $l=0,1 \dots 60$. Note, this search scale is not required in a realistic control scenario. To convert the strict inequality Equation \eqref{eq:strictinequality} into a non-strict inequality Equation \eqref{eq:nonstrictinequality}, $\gamma=0.001$ has been used. In each model the objective function to be minimized was the linear cost function of power generation, where $\epsilon_g$ in Table \ref{tab:generators} were the cost coefficients. Apart from the disjunctive constraints, the objective function was subject to all node balances and line flow constraints of the DC approximation. For this second offline parameter search, $100$ operating points were sampled from the correlated load distribution and for all $60$ models, $100$ MILPs (in total $6000$) were solved to compute the output of the control approach (dispatch decisions). Finally, the true label was evaluated based on the dispatch decisions. All MILPs were implemented using Pyomo 5.1.1 \cite{Har17} and Gurobi 7.02 \cite{Gur16} was used as a solver. 

\vspace{-0.5em}
\subsection{Computational complexity}
\vspace{-0.5em}
For the online computational complexity, the computation time of solving one single MILP with the proposed disjunctive approach was on average $\SI{26}{\percent}$ lower than to solve all $82$ LPs with separate constraints from the terminal nodes. The main computational benefit results from training the safety margin $\alpha$ offline and avoiding the online computation of the labels (e.g., with dynamic simulations as in \cite{Xu14}). 

\vspace{-0.5em}
\subsection{Rule accuracy}
\vspace{-0.5em}
For the rule accuracy, Fig. \ref{fig:outputf1} shows the mismatch of control error and the test error based on the training population. For instance for $\alpha_1=0$, the average test error was $\leq \SI{0.02}{\percent}$, thus the DT was almost perfect as predictor. However, when the rules are used for control, the actual average control error was $\geq \SI{70}{\percent}$. The figure also shows the ability of the proposed method to reach $\SI{100}{\percent}$ operating points corresponding to the acceptable class by increasing $\alpha$. Specifically, by adding $\alpha \geq \SI{48}{\MW}$ to the rule, acceptable operation could be guaranteed for correlated $\pm \SI{25}{\percent}$ uncertainty in all loads. 

\vspace{-0.5em}
\subsection{Cost-effectiveness}
\begin{figure}
\centering
\includegraphics[scale=0.47]{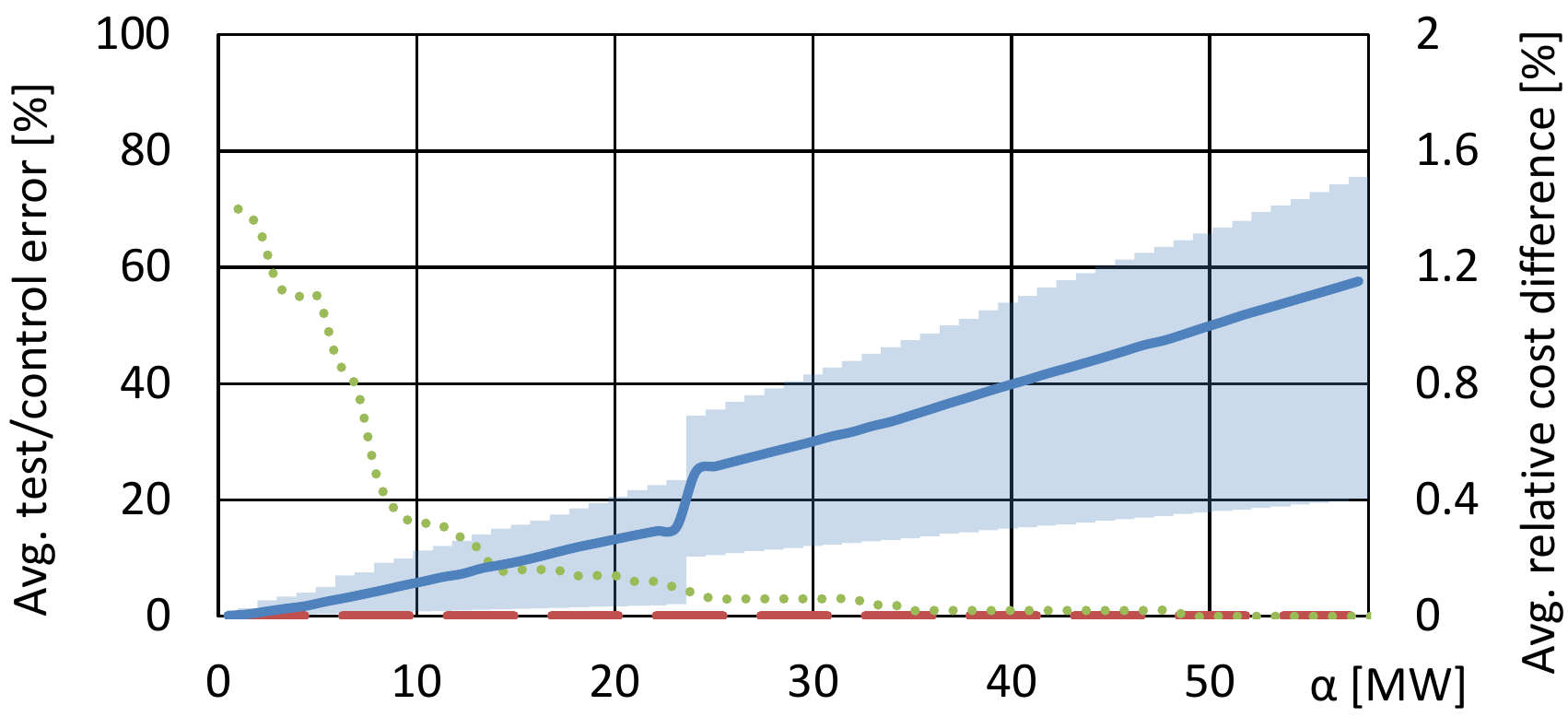}
\caption{ \footnotesize Exhaustive search to study the safety margin $\alpha$. The DT average test error and the average control error are shown as a red dashed and a green dotted line, respectively. The average relative cost difference to the SCOPF solution and the $10$th and $90$th percentiles are shown in blue.}
\label{fig:outputf1}\vspace{-1.5em}
\end{figure}

\vspace{-0.5em}
For all $100$ samples, the cost of the proposed sample-derived disjunctive approach was compared against the SCOPF solution. The average relative cost difference (in blue) is presented in Fig. \ref{fig:outputf1}. Note that only operating points that lead to an acceptable solution are used for this comparison.
It can be seen that the economic cost difference to the SCOPF solution was generally small (relative difference is $\leq \SI{0.003}{\percent}$ for $\alpha_0=\SI{0}{\MW}$ and roughly $\SI{1}{\percent}$ for $\alpha=\SI{48}{\MW}$). This was because all cost coefficients are in a very narrow range (Table \ref{tab:generators}). The discontinuous jump around $\alpha =\SI{23}{\MW}$ resulted from a particular cost-effective terminal node that was excluded for $\alpha \geq \SI{23}{\MW}$. For $\alpha=48$, where the operator could guarantee acceptable operation under all potential faults for the complete uncertainty spectrum, the average Euclidean distance of the generator powers to the SCOPF dispatch solution was $\SI{229}{\MW}$ that was $\SI{3.8}{\percent}$ of the average total power.

\vspace{-0.5em}
\section{Conclusion}\label{sec:conc}
\vspace{-0.5em}
We presented the specific challenges when sample-derived rules are embedded in optimal decision--making for control of a power system. Particularly challenging are the computational performance and accuracy of rules. We have introduced a novel disjunctive approach to deal with the computational challenge and a grid search strategy using a safety margin to deal with the computational challenge and the inaccuracy of rules. We studied the challenges and the solution strategies by using the IEEE 39 bus network. The proposed disjunctive approach is able to secure system's dispatch decision against a user-specified (stability-) criterion for a wide range of operational uncertainty. In a steady-state comparison, the approach resulted in $\SI{1}{\percent}$ higher costs than an oracle model and the online-computational cost is low since online simulations are avoided. Moreover, the proposed sample-derived disjunctive approach provides a framework capable of accommodating a wide variety of linear and ensemble classifiers.
In future works, the implementation of ensembles could be studied and reduction in operational and online computational cost could be achieved by learning safety margins individually for the terminal nodes instead of a generalized margin. Further offline computations could be reduced by using importance sampling techniques.

\bibliographystyle{IEEEtran}
\bibliography{MyLib}

\begin{thebibliography}{10}
\providecommand{\url}[1]{#1}
\csname url@samestyle\endcsname
\providecommand{\newblock}{\relax}
\providecommand{\bibinfo}[2]{#2}
\providecommand{\BIBentrySTDinterwordspacing}{\spaceskip=0pt\relax}
\providecommand{\BIBentryALTinterwordstretchfactor}{4}
\providecommand{\BIBentryALTinterwordspacing}{\spaceskip=\fontdimen2\font plus
\BIBentryALTinterwordstretchfactor\fontdimen3\font minus
  \fontdimen4\font\relax}
\providecommand{\BIBforeignlanguage}[2]{{%
\expandafter\ifx\csname l@#1\endcsname\relax
\typeout{** WARNING: IEEEtran.bst: No hyphenation pattern has been}%
\typeout{** loaded for the language `#1'. Using the pattern for}%
\typeout{** the default language instead.}%
\else
\language=\csname l@#1\endcsname
\fi
#2}}
\providecommand{\BIBdecl}{\relax}
\BIBdecl

\bibitem{Weh98}
L.~A. Wehenkel, \emph{Automatic Learning Techniques in Power Systems}.\hskip
  1em plus 0.5em minus 0.4em\relax Kluwer Academic Publishers, 1998.

\bibitem{Kri11}
V.~Krishnan, J.~D. McCalley, S.~Henry, and S.~Issad, ``Efficient database
  generation for decision tree based power system security assessment,''
  \emph{IEEE Transactions on Power Systems}, vol.~26, no.~4, pp. 2319--2327,
  2011.

\bibitem{Kon16}
I.~Konstantelos, G.~Jamgotchian, S.~Tindemans, P.~Duchesne, S.~Cole, C.~Merckx,
  G.~Strbac, and P.~Panciatici, ``Implementation of a massively parallel
  dynamic security assessment platform for large-scale grids,'' \emph{IEEE
  Transactions on Smart Grid}, vol.~PP, no.~99, pp. 1--1, 2016.

\bibitem{Kar02}
E.~S. Karapidakis and N.~D. Hatziargyriou, ``Online preventive dynamic security
  of isolated power systems using decision trees,'' \emph{IEEE Transactions on
  Power Systems}, vol.~17, no.~2, pp. 297--304, 2002.

\bibitem{Xu14}
Y.~Xu, Z.~Y. Dong, R.~Zhang, and K.~Po~Wong, ``A decision tree-based on-line
  preventive control strategy for power system transient instability
  prevention,'' \emph{International Journal of Systems Science}, vol.~45,
  no.~2, pp. 176--186, 2014.

\bibitem{Gen10}
I.~Genc, R.~Diao, V.~Vittal, S.~Kolluri, and S.~Mandal, ``Decision tree-based
  preventive and corrective control applications for dynamic security
  enhancement in power systems,'' \emph{IEEE Transactions on Power Systems},
  vol.~25, no.~3, pp. 1611--1619, 2010.

\bibitem{Cos16}
D.~C.~L. Costa, M.~V.~A. Nunes, J.~P.~A. Vieira, and U.~H. Bezerra, ``Decision
  tree-based security dispatch application in integrated electric power and
  natural-gas networks,'' \emph{Electric Power Systems Research}, vol. 141, pp.
  442--449, 2016.

\bibitem{Tha17}
F.~Thams, L.~Halilbašic, P.~Pinson, S.~Chatzivasileiadis, and R.~Eriksson,
  ``Data-driven security-constrained opf,'' in \emph{10th Bulk Power Systems
  Dynamics and Control Symposium}, 2017, Conference Proceedings.

\bibitem{Liu14}
C.~Liu, K.~Sun, Z.~H. Rather, Z.~Chen, C.~L. Bak, P.~Thøgersen, and P.~Lund,
  ``A systematic approach for dynamic security assessment and the corresponding
  preventive control scheme based on decision trees,'' \emph{IEEE Transactions
  on Power Systems}, vol.~29, no.~2, pp. 717--730, 2014.

\bibitem{Gro94}
R.~Raman and I.~E. Grossmann, ``Modelling and computational techniques for
  logic based integer programming,'' \emph{Computers \& Chemical Engineering},
  vol.~18, no.~7, pp. 563--578, 1994.

\bibitem{Jam14}
G.~James, D.~Witten, T.~Hastie, and R.~Tibshirani, \emph{An Introduction to
  Statistical Learning: with Applications in R}.\hskip 1em plus 0.5em minus
  0.4em\relax Springer Publishing Company, Incorporated, 2014.

\bibitem{Bre84}
L.~Breiman, J.~H. Friedman, R.~A. Olshen, and C.~J. Stone, ``Classification and
  regression trees,'' \emph{Wadsworth \& Brooks Monterey, CA}, 1984.

\bibitem{Bal85}
E.~Balas, ``Disjunctive programming and a hierarchy of relaxations for discrete
  optimization problems,'' \emph{SIAM Journal on Algebraic Discrete Methods},
  vol.~6, no.~3, pp. 466--486, 1985.

\bibitem{Ana12}
A.~Pai, \emph{Energy function analysis for power system stability}.\hskip 1em
  plus 0.5em minus 0.4em\relax Springer Science \& Business Media, 2012.

\bibitem{Ped11}
F.~Pedregosa, G.~Varoquaux, A.~Gramfort, V.~Michel, B.~Thirion, O.~Grisel,
  M.~Blondel, P.~Prettenhofer, R.~Weiss, and V.~Dubourg, ``Scikit-learn:
  Machine learning in python,'' \emph{Journal of Machine Learning Research},
  vol.~12, no. Oct, pp. 2825--2830, 2011.

\bibitem{Har17}
W.~E. Hart, C.~D. Laird, J.-P. Watson, D.~L. Woodruff, G.~A. Hackebeil, B.~L.
  Nicholson, and J.~D. Siirola, \emph{Pyomo-optimization modeling in
  python}.\hskip 1em plus 0.5em minus 0.4em\relax Springer Science \& Business
  Media, 2017, vol.~67.

\bibitem{Gur16}
I.~Gurobi~Optimization, ``Gurobi optimizer reference manual,'' 2016.

\end{thebibliography}
\end{document}